\documentclass[prd,
12pt,twocolumn, groupedaddress, preprintnumbers,nofootinbib,floatfix]{revtex4-2}
\usepackage[top=3cm, bottom=2.1cm, left=2cm, right=2cm]{geometry}
\pdfoutput=1

\usepackage{comment}

\usepackage{graphicx}
\usepackage{url}
\usepackage[bookmarks, pagebackref=false]{hyperref}
\usepackage[usenames,dvipsnames]{xcolor}
\usepackage{dcolumn}
\usepackage{bm}
\usepackage{bbm}
\usepackage{amsmath,amssymb,amsfonts}
\usepackage{color}
\usepackage{hyperref}
\usepackage[Symbolsmallscale]{upgreek}
\usepackage{dsfont}
\usepackage[all]{xy}
\usepackage{pstricks}
\usepackage{dsfont}%
\usepackage{mathtools}
\usepackage{placeins}
\usepackage{enumerate}
\usepackage{cases}
\usepackage{xspace}
\usepackage{bbold}
\usepackage{cancel} 
\usepackage{slashed}
\usepackage{subfigure}
\usepackage{natbib}
\usepackage{ulem}

\newcommand{\be}{\begin{equation}}
	\newcommand{\ee}{\end{equation}}
\newcommand{\beq}{\begin{eqnarray}}
	\newcommand{\eeq}{\end{eqnarray}}

\begin{document}

\title{Identification of the low energy excess in dark matter searches with crystal defects}

\author{Matti Heikinheimo}
\email{matti.heikinheimo@helsinki.fi}
\affiliation{Department of Physics, University of Helsinki, 
                      P.O.Box 64, FI-00014 University of Helsinki, Finland}
\affiliation{Helsinki Institute of Physics, 
                      P.O.Box 64, FI-00014 University of Helsinki, Finland}
                      
\author{Sebastian Sassi}
\email{sebastian.k.sassi@helsinki.fi}
\affiliation{Department of Physics, University of Helsinki, 
                      P.O.Box 64, FI-00014 University of Helsinki, Finland}
\affiliation{Helsinki Institute of Physics, 
                      P.O.Box 64, FI-00014 University of Helsinki, Finland}

\author{Kai Nordlund}
\email{kai.nordlund@helsinki.fi}
\affiliation{Department of Physics, University of Helsinki, 
                      P.O.Box 64, FI-00014 University of Helsinki, Finland}
\affiliation{Helsinki Institute of Physics, 
                      P.O.Box 64, FI-00014 University of Helsinki, Finland}

\author{Kimmo Tuominen}
\email{kimmo.i.tuominen@helsinki.fi}
\affiliation{Department of Physics, University of Helsinki, 
                      P.O.Box 64, FI-00014 University of Helsinki, Finland}
\affiliation{Helsinki Institute of Physics, 
                      P.O.Box 64, FI-00014 University of Helsinki, Finland}

\author{Nader Mirabolfathi}                      
\email{mirabolfathi@physics.tamu.edu} 
\affiliation{Department of Physics and Astronomy and the Mitchell Institute for Fundamental Physics and Astronomy,
Texas A\&M University, College Station, TX 77843, USA}


\begin{abstract}
\noindent
An excess of events of unknown origin at low energies below 1 keV has been observed in multiple low-threshold dark matter detectors. Understanding the origin of these events is of utmost importance, as this unidentified event rate currently overwhelms any potential new physics signal. Depending on the target material, nuclear recoil events at these energies may cause lattice defects, in which case a part of the true recoil energy is stored in the defect and not observed in the phonon detector. If the threshold for defect creation is sharp, this effect leads to a prominent feature in the observed recoil spectrum. Electronic recoils at low energies do not create defects and therefore the feature in the observed spectrum is not expected in that case. We propose to use the sharp defect creation threshold of diamond to test if the low energy events are due to nuclear recoils. Based on simulated data we expect the nuclear recoil peak in the observed spectrum to be visible in diamond with a data set of $\sim700$ events, potentially achievable with $\sim0.1$ gram days of exposure.
\end{abstract}
\preprint{HIP-2021-51 /TH}
\maketitle


\section{Introduction}

The possible existence of weakly interacting massive particles (WIMP) is a well motivated paradigm to address 
the cosmological and astrophysical observations~\cite{Planck:2018vyg, Bertone:2004pz}. On one hand this paradigm allows for an explanation of how the dark matter abundance arises during the thermal evolution of the early universe and, consequently, leads to expectation of directly detectable interactions between WIMPs and baryonic matter~\cite{Arcadi:2017kky, Roszkowski:2017nbc}. Many experiments have searched 
for such interactions~\cite{Schumann:2019eaa}, with most stringent exclusion constraints~\cite{XENON:2018voc} applying to WIMPs with masses above ${\cal O}$(10 GeV).

Progress into the sub-GeV region has been demonstrated by new detector developments offering detection thresholds sensitive to WIMP-nucleus scattering at precision of single-electron excitation~\cite{Romani:2017iwi,Crisler:2018gci,DAMIC:2020cut,EDELWEISS:2020fxc}. While these technologies are currently operating with low mass targets, ${\cal O}$(g), attempts to scale up to detector masses of ${\cal O}$(kg) have been made~\cite{Mirabolfathi:2015pha,Iyer:2020nxe}. Phonon-mediated detectors with ${\cal O}$(eV) resolution are the most appropriate detectors to perform the measurements that we are proposing in this work. Phonons are among the lowest energy quantum excitations (compred to \textit{e.g.} ionization and scintillation) that can be detected after particle interactions. In addition to their excellent signal to noise, phonon-mediated detectors offer an interaction-type independent (nuclear or electron recoil) energy measurement .
Many groups have recently achieved the energy resolution that are within the required range for our proposed detection technique \cite{Strauss:2017cam, CPD:2020xvi,Verma:2022tkq}. 

Recently several experiments \cite{CRESST:2019axx,CRESST:2019jnq,DAMIC:2020cut,EDELWEISS:2019vjv,EDELWEISS:2020fxc,CRESST:2017ues,NUCLEUS:2019kxv,SENSEI:2020dpa,SuperCDMS:2018mne,SuperCDMS:2020ymb} have observed a steeply rising event rate at low energies, $E_{\rm r} \lesssim 1000$ eV. The origin of these events is currently unknown, and understanding their physical character is a question of great interest for both the DM and coherent neutrino scattering experiments~\cite{Proceedings:2022hmu}. Most of the anticipated background sources, such as photons or electrons, would give rise to electron recoils. Therefore the identification of the nuclear/electron recoil character of these events would add an important piece of information towards understanding and mitigating this background. 

We propose to approach this problem with detailed understanding of the response of the target material to low energy scattering events, achieved via molecular dynamics (MD) simulations~\cite{Gib60,Dia87,Nor18b,PhysRevB.78.045202,NORDLUND2006322,Sassi:2022njl}. Particularly, it was noticed in \cite{Kadribasic:2020pwx,Sassi:2022njl} that a sharp defect creation threshold in diamond leads to a prominent feature in the observed nuclear recoil spectrum: the energy readout of the phonon detector will not see the energy stored in the defect, and the affected recoil events will be shifted towards lower observed energies. If the defect creation threshold is sharp, this effect turns on abruptly at the corresponding recoil energy, resulting in a prominent peak followed by a dip in the observed event rate. On the other hand, this effect is not expected for electronic recoils, and therefore its presence or absence in the recoil spectrum allows to discriminate between these two explanations for the unknown origin of the events. Beyond particle scattering events, electrical noise and stress relaxation or microscopic fractures in the material have been suggested as possible sources of the excess events. Electrical noise would clearly not be affected by energy loss and therefore the observation of this effect would also exclude this explanation. While we do not expect the stress relaxation/fracture scenario to exhibit the same feature either, a full understanding of the expected spectrum in this case requires dedicated simulations and is beyond the scope of this work.

Phonon-mediated detectors have the advantage of fully measuring the recoil energy without making any assumption about the nature of the recoil: nuclear or electronic. Recent progress in phonon-mediated detectors allows for the low detection thresholds that are required to measure these lattice defect features~\cite{CPD:2020xvi,EDELWEISS:2019vjv,Strauss:2017cam}. In particular, gram scale diamond based detectors are expected to offer a resolution that is superior to the existing technologies~\cite{Kurinsky:2019pgb,Abdelhameed:2022skh}. 
In the following, we will describe this effect in more detail in light of 
recent experimental results and propose a method to assess the underlying nature of the observed low energy event rate.\\

\section{Testing the nuclear recoil origin of the excess}
\begin{figure*}[htb]
    \begin{center}
		\includegraphics[width=0.32\linewidth]{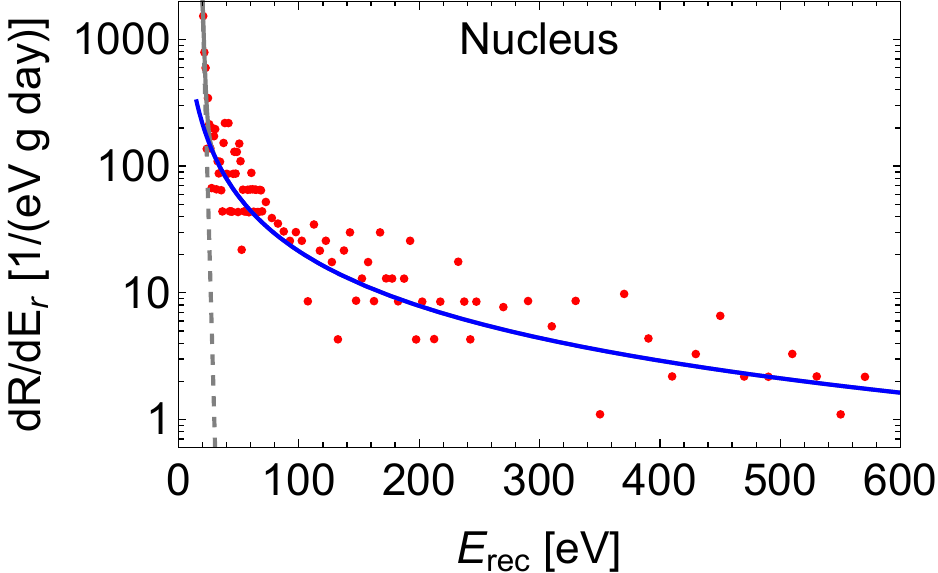}
		\includegraphics[width=0.32\linewidth]{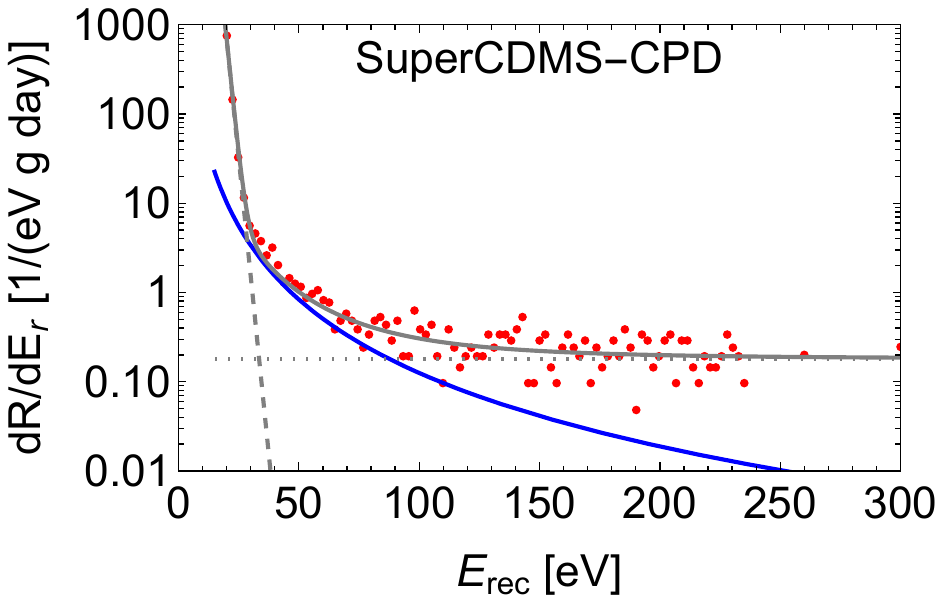}
		\includegraphics[width=0.32\linewidth]{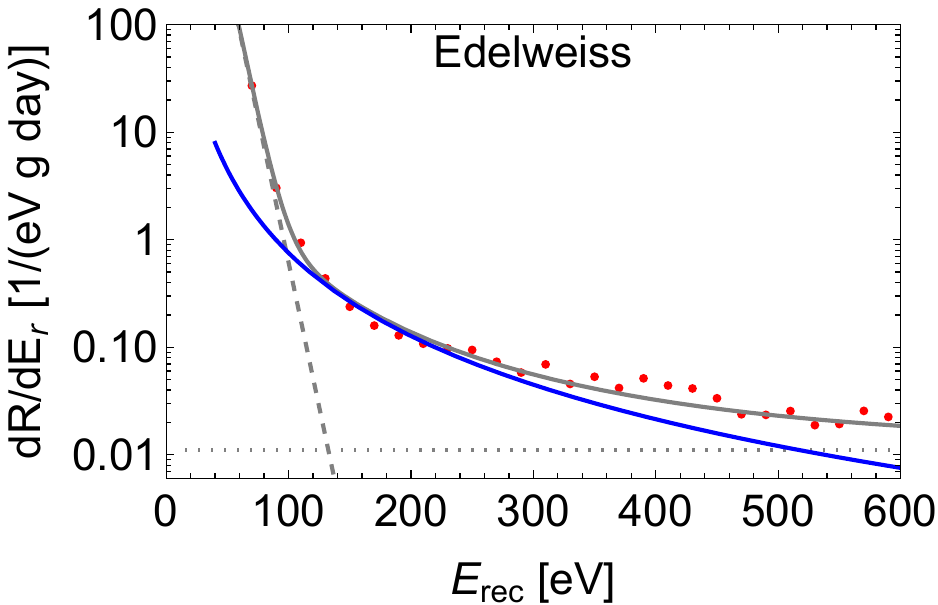}
		\caption{The differential event rate $dR/dE_{\rm r}$ observed in the Nucleus 1g-prototype (left), SuperCDMS-CPD (center) and Edelweiss (right) measurements shown with the red points. The fit function (\ref{nucleusfit}) is shown with the gray solid line. The gray dashed line shows the exponential component, the gray dotted line the constant component and the blue solid line the power-law component of the fit function, taken here to represent the unidentified excess. }
		\label{nucleusdata}
    \end{center}
\end{figure*}

To test if the excess events are due to nuclear recoils, we propose to use a diamond detector. 
For a quantitative analysis, we use data from Nucleus 1g prototype \cite{NUCLEUS:2019kxv}, SuperCDMS-CPD \cite{SuperCDMS:2020aus} and Edelweiss \cite{EDELWEISS:2019vjv} shown in figure \ref{nucleusdata}, reproduced here using the data repository \cite{excess_workshop_2021}. We will parametrize this data using a three-component fit, with an exponential, a power law and a constant component, of the form
\be
f(x) = Ae^{-\alpha x}+Bx^\beta+C,
\label{nucleusfit}
\ee
where $x= E_{\rm r}/{\rm eV}$ and we have determined the best fit values for the parameters as shown in table \ref{bestfitparams}. The fit function is shown together with the data points in figure \ref{nucleusdata}. A similar fit for the SuperCDMS and Edelweiss data was used in \cite{Abbamonte:2022rfh}, where it was suggested that the exponential part is due to electronic trigger noise. We adopt this interpretation, and therefore do not apply the energy loss effect to the exponential part of the event rate in our simulations, as these counts are taken to not represent real recoil events. Furthermore the constant component $C$ is taken to represent background from mostly electron recoils, so that the power law component represents the unidentified excess events. Therefore in our simulations the energy loss is not applied to the constant component either, but we have checked that the results do not significantly differ based on this decision. 

\begin{table*}[htb]
    \centering
    \begin{tabular}{c|c c c c c}
          & $A$  & $\alpha$ & $B$  & $\beta$ & $C$ \\
        \hline
        Nucleus & $(9.7\pm 25.7)\times10^9$\, & $0.77\pm 0.13$\, & $(1.58\pm 0.40)\times 10^4$\, & $-1.44\pm0.05$\, & $0\pm 0.19$ \\
        SuperCDMS & $(1.41\pm 0.16)\times 10^8$ & $0.61\pm 0.006$ & $(3.7\pm 4.1)\times 10^4$ & $-2.7\pm 0.3$ & $0.18\pm 0.01$\\
        Edelweiss & $(1.46\pm 0.28)\times 10^5$ & $0.124\pm 0.003$ & $(1.04\pm 0.55)\times 10^5$ & $-2.6\pm 0.1$ & $0.011\pm 0.002$
    \end{tabular}
    \caption{Best fit values for the parametric model (\ref{nucleusfit}) for the three data sets. The parameters $A,B,C$ are in units of events/(eV g day). }
    \label{bestfitparams}
\end{table*}

\begin{figure}[h!]
    \begin{center}
		\includegraphics[width=0.99\linewidth]{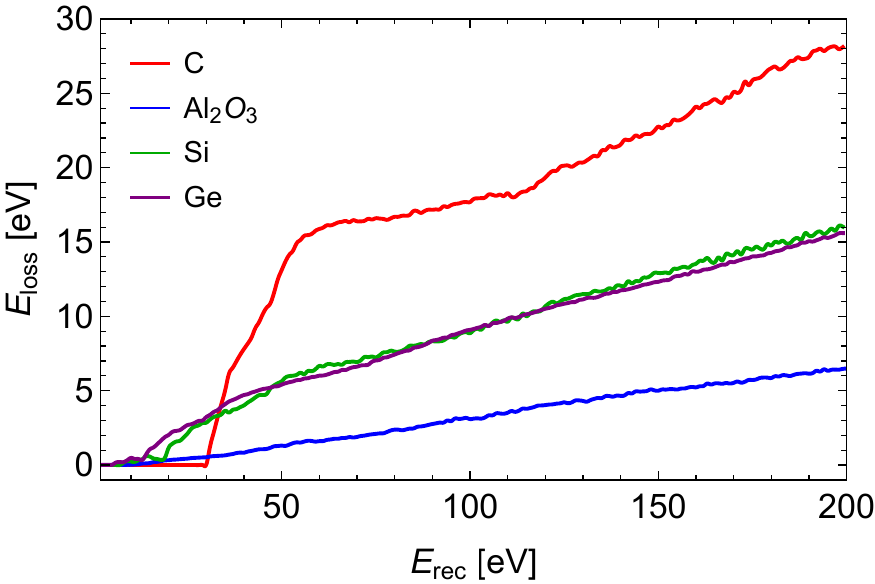}
		\caption{The average (over recoil direction) energy loss in diamond (red), sapphire (blue), silicon (green) and germanium (purple) as a function of the recoil energy.}
		\label{ElossAverageC}
    \end{center}
\end{figure}

Based on MD simulation data presented in~\cite{Sassi:2022njl}, diamond has a very sharp threshold for defect creation, resulting in a step-like rise in the average energy loss as a function of recoil energy, shown in figure \ref{ElossAverageC}. As the energy read out from the phonon detector will not see the energy stored in the defects, this sharp onset of energy loss will result in a peak in the visible energy spectrum, if the underlying recoil spectrum is smooth. If the events are electronic recoils, no energy loss is expected and the phonon measurement should see the smooth underlying spectrum. For comparison, figure \ref{ElossAverageC} shows also the average energy loss curves for silicon in green and germanium in purple. For these semiconductor targets we can observe a threshold, but smoother than in diamond and at a lower energy. Silicon and germanium have identical crystal structure and very similar chemical properties, which explains the similar energy loss curves of these materials.

The blue line in figure \ref{ElossAverageC} shows the average energy loss in sapphire, based on~\cite{Sassi:2022njl} . Evidently for sapphire the energy loss is a rather smooth function of recoil energy and therefore a peak is not expected in a sapphire detector, such as the Nucleus 1g. We can therefore expect that if the low energy events observed by Nucleus are nuclear recoil events, the Nucleus data represents the true underlying recoil spectrum and a diamond detector should see a spectrum that contains a peak due to the energy loss feature. As the thresholds in germanium and silicon are modest, the same conclusion holds to a large degree also for SuperCDMS (Si) and Edelweiss (Ge).
 
 \begin{figure*}[h!tb]
    \begin{center}
    \includegraphics[width=\linewidth]{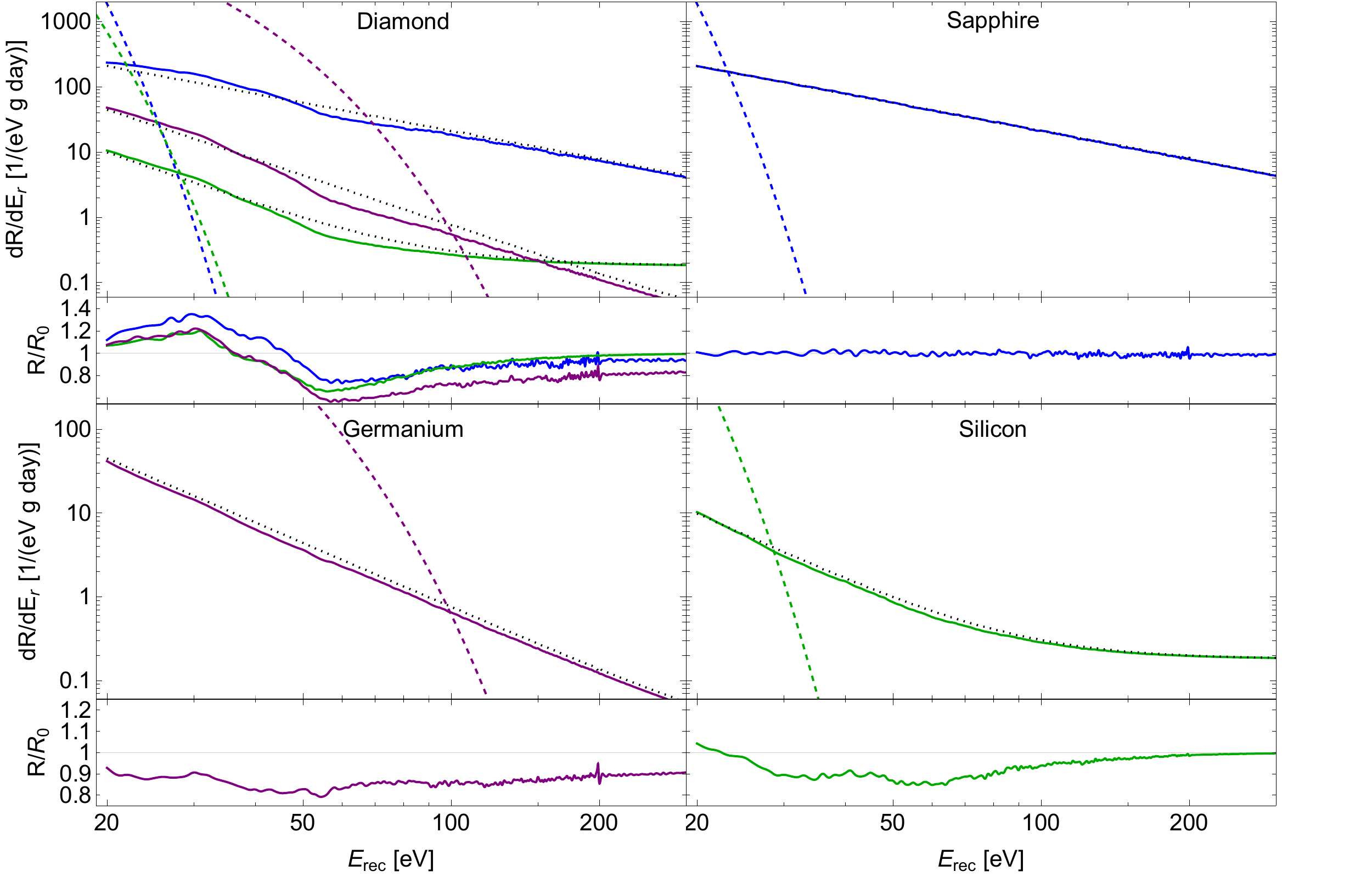}
		\caption{Top left: The observed event rate in diamond for the underlying recoil spectrum given by the power law component of the fit function  (\ref{nucleusfit}). The blue lines correspond to the best fit parameter values for Nucleus data, the green lines for SuperCDMS data and the purple lines for Edelweiss data. The solid curves show the observed rate after the energy loss, and black dotted curves if the energy loss is not simulated. The dashed curves show the exponential component of each fit. The curve in the bottom inset shows the ratio of the rate with/without the energy loss for the power law component. Top right: The same for a sapphire detector, using Nucleus fit. The bottom row shows same results for germanium (left, Edelweiss fit) and silicon (right, SuperCDMS fit).}
		\label{dRdEplots}
    \end{center}
\end{figure*}

 To confirm this hypothesis, we have taken the power law part of the fit function (\ref{nucleusfit}) to represent the true underlying excess event rate, over which the scatter in the data is assumed to be statistical fluctuations, and simulated the resulting observed energy spectrum in diamond, sapphire, germanium and silicon detectors, assuming 1 eV energy resolution with Gaussian smearing. Due to the unknown origin of the events, we assume the same normalization for the event rate in units of events/[eV gram day] in each material, not correcting for possible differences e.g. in the nuclear scattering cross sections for different nuclei. While our sample data from the three experiments utilizing different target materials show clearly distinct event rates, we do not determine in detail to what part of the difference in overall normalization is explained e.g. by differences in the shielding or detector efficiencies of the experiments, and which part is due to underlying scattering cross sections. 
 
 The simulation is performed using 1 eV energy intervals. We sample recoil directions randomly, and for each direction increase the energy in 1 eV intervals. For each (direction,energy) combination we obtain energy loss due to defect creation by comparing the potential energy given by the MD interatomic potentials before and after the recoil event. Details of how the MD simulations are set up are provided in Refs. \cite{Sassi:2022njl,Kad17,Kadribasic:2020pwx}. This energy is then subtracted from the true recoil energy to produce the observed energy in the phonon measurement. The resulting observed recoil spectra are shown in figure \ref{dRdEplots} with the solid lines, while the black dotted lines show the observed event rate if the energy loss is not included in the simulation.

 The top left panel shows the result for diamond and the top right panel for a sapphire detector. Indeed, for sapphire the solid and dotted lines are indistinguishable, confirming that we can treat the sapphire data as representing the true underlying event rate. For diamond we should instead expect to see a peak followed by a dip over the smooth power law. If the events are not due to nuclear recoils, this feature will not appear in the observed spectrum in diamond. Therefore the presence of the peak can be used as a verification of the nuclear recoil origin of the events. 
 
 In the bottom row of figure \ref{dRdEplots} we repeat this calculation for germanium, shown in the left panel, and for silicon, shown on the right panel. For these materials the defect creation threshold is in the 10--20 eV range, and therefore lies outside the selection window of our current data sample, and would in any case be masked by the exponential trigger noise. The effect of the energy loss is then just to move the expected curve slightly towards lower energy, but no prominent peak is produced such as in diamond. These considerations favor diamond as the best suited material for this task, where the threshold is both sharp and appears at high enough recoil energy, so that the peak in the observed spectrum appears over the smooth power-law component in our event rate model.

\section{Statistical analysis}
To quantify the statistical significance of the energy loss effect in a diamond detector, we generated simulated data sets from the expected event rate in diamond for the three best fit parameter sets shown in table \ref{bestfitparams}, and accounting for the energy loss as described above. We then computed the log-likelihood ratio for fitting the simulated data with the event rate containing the energy loss effect, allowing the overall normalization of the event rate vary but keeping the fit parameters $\{A,\alpha,B,\beta,C\}$ fixed to their best fit values, and with an event rate assuming no energy loss, but allowing to vary the fit parameters. This method allows to test if the effect of the energy loss could be mimicked by altering the fit parameters, which could mask the effect as the values of the parameters are not a priori known, and appear also to vary between experiments utilizing different target materials. Because we assume that the power law component of the fit function represents the excess part, we have repeated this procedure also for the fit function containing just the power law component, (i.e. omitting the exponential and constant components), and for a fit containing the power law and the constant components but omitting the exponential. The corresponding test statistic is given by
\be
q_0 = 2\log \left( \frac{{\rm max}\, \mathcal{L}(\mu_{\rm loss})}{ {\rm max}\, \mathcal{L}(A,\alpha,B,\beta,C)}  \right),
\ee
where $\mathcal{L}(\mu_{\rm loss})$ and  $\mathcal{L}(A,\alpha,B,\beta,C)$ are the likelihoods for drawing the data from the expected distribution of the event rate with or without the energy loss effect, respectively given by
\be
\mathcal{L}(\{\lambda\}) = \prod\limits_{i=1}^N \frac{ e^{-  n_{{\rm exp},i}(\{\lambda\})}}{n_{{\rm obs},i}!}\left( n_{{\rm exp},i}(\{\lambda\}) \right)^{n_{{\rm obs},i}}.
\ee
Here $\{\lambda\} = \{ \mu_{\rm loss}\},\{ A,\alpha,B,\beta,C\}$ are the parameters that are varied to find the maximum likelihood, $n_{{\rm exp},i}(\{\lambda\})$ is the expected number of events in the energy bin $i$ for the event rate, and $n_{{\rm obs},i}$ is the 'observed' number of events in the bin $i$ in the simulated data set. For the number of bins we use $N=180$ corresponding to 1 eV bins from 20 eV to 200 eV.

For varying exposure we simulate 1000 data sets, and test if $q_0 > 9$ in at least 90\% of the iterations. If this test is successful, we conclude that the corresponding exposure is enough to identify the events as nuclear recoils at $3\sigma$ confidence level. The results of this analysis are shown in table \ref{tab:Results}, where we have used three versions of the event rate model: {\it{(i)}} the full model (\ref{nucleusfit}), {\it{(ii)}} the model without the exponential component and {\it{(iii)}} the model with only the power law component.

\begin{table*}[htb]
    \centering
    \begin{tabular}{c|c c | c c | c c}
      & \multicolumn{2}{c}{Full fit} & \multicolumn{2}{c}{Power law + const} & \multicolumn{2}{c}{Power law only} \\
      & $\mathcal{E}$ [gd] & $N_{\rm events}$  & $\mathcal{E}$ [gd]  & $N_{\rm events}$  & $\mathcal{E}$ [gd]  & $N_{\rm events}$ \\
    \hline
    Nucleus & 0.08 & 700  & 0.11  & 710 & 0.11  & 710 \\
    SuperCDMS  & 6.3  & 7\,900 & 17  & 2\,500 & 3.8  &  440  \\
    Edelweiss  & 750 & 190\,000 & 2.3  &  1\,300 & 0.75  & 440 
    \end{tabular}
    \caption{Required exposure in gram days and the corresponding number of events for a $3\sigma$ observation of the energy loss effect in the recoil spectrum, for the three best fit parameter sets from table \ref{bestfitparams}, using the full fit function, omitting the exponential component, and omitting the exponential and constant components.}
    \label{tab:Results}
\end{table*}

Looking at the first column, we notice that the larger amount of required events for SuperCDMS and especially for the Edelweiss fits is explained by the fact that the exponential part dominates the event rate up to higher energy than in the Nucleus fit, partly masking the peak feature, as shown by the dashed lines in the top left panel of figure \ref{dRdEplots}. These numbers point to the importance of suppressing the trigger noise so that it does not mask the event rate in the relevant window above $\gtrsim 20$ eV in order to make full use of the energy loss feature.

The second column shows the results for the fit function where the exponential part has been omitted; Assuming that the exponential component represents trigger noise or other measuring electronics related background, this will be device specific and not necessarily present in similar magnitude in the proposed diamond detector. Therefore it is interesting to examine how the event rate would look like in the absence of this noise component. For the Nucleus fit, we find a similar number of events required for the observation of the energy loss effect, as was the case with the exponential component present. This confirms the conclusion that if the exponential rise begins only at energies sufficiently below the defect creation threshold, as is the case for this fit, the peak feature due to energy loss will not be masked by it and the significance of the observation therefore does not depend on the presence or absence of this background. For the other two fits we instead find a clear reduction in the required number of events compared to the first column, and as expected the effect is more dramatic for the Edelweiss fit where the exponential masked a larger part of the recoil spectrum.

Finally, the third column presents the results for the pure power law model, omitting also the constant component. Since the Nucleus best fit value for the constant is zero, these numbers are identical to the second column for the Nucleus parameters. For the SuperCDMS and Edelweiss fits the power law indices are almost identical, resulting in equal numbers of required events. It appears that the steeper power law $\sim 2.7$ of SuperCDMS/Edelweiss fits compared to the \mbox{$\sim 1.4$} of the Nucleus fit makes the observation of the energy loss effect easier, with the required number of events 440 against 710 for the Nucleus fit.

\section{Conclusions}
The energy spectrum of the unidentified excess events observed at low energy can be reasonably fitted with a power law model. Assuming this smooth function as the true underlying event rate, we have shown that the nuclear recoil origin of these events can be tested with a diamond detector, due to the sharp defect creation threshold which gives rise to a prominent peak in the observed recoil spectrum. If the trigger noise can be suppressed down to $\sim\! 20$ eV this peak can be observed in a data set of $\sim\! 700$ events in the selection window [20-200]~eV with a diamond detector with 1~eV energy resolution. Assuming an event rate similar to the one observed in Nucleus, this corresponds to $\sim\!0.1$ gram days of data. In sapphire, germanium and silicon no peak is expected due to the smooth energy loss function, so that the data from existing measurements can reasonably be used as reference spectrum.

There is an ongoing research to calibrate ionization yield for very low energy nuclear recoils in Si and Ge detectors using low energy neutron beam and simultaneous scattering angle measurement\footnote{See e.g. presentation by Tarek Saab at the Excess2022 workshop, \url{https://indico.scc.kit.edu/event/2575/contributions/9684/}}. Very similar method can be used to validate and calibrate the effects of crystalline defects in nuclear recoil energy spectrum in our proposed diamond phonon mediated detectors.

\section*{Acknowledgements}
We thank Jesper Byggmästar and Antti Kuronen for their help in setting up the MD simulations, the Finnish Computing Competence Infrastructure (FCCI) for supporting this project with computational and data storage resources, and the Academy of Finland for financial support under project $\# 342777$.


\bibliography{main.bib}
\bibliographystyle{hieeetr}	

	
\end{document}